\documentclass[11pt]{revtex4-1}

\usepackage{graphicx}

\begin{document}

\title{ A Dynamical System Analysis of Holographic Dark Energy Models with Different IR Cutoff}
%\author{Nilanjana Mahata\footnote {nmahata@math.jdvu.ac.in}}

\author{Nilanjana Mahata\footnote {nmahata@math.jdvu.ac.in}}
\author{Subenoy Chakraborty\footnote {schakraborty@math.jdvu.ac.in}}
\affiliation{Department of Mathematics, Jadavpur University, Kolkata-700032, West Bengal, India.}

%%%%%%%%%%%%%%%%%%%%%%%%%%%%%%%%%%%%%%%%%%%%%%%%%%%%%%%%%%%%%%%%%%%%%%%%%%%%%%%%%%%%%%%%%%%%%%%%%%%%%%%%%%%%%%%%%%%%%%%%

%\date{\today}

\begin{abstract}
The paper deals with a dynamical system analysis of the cosmological evolution of an   holographic  dark energy (HDE) model interacting with dark matter (DM) which  is chosen in the form of dust. The infrared cut-off of the holographic model is considered as  future event horizon or Ricci length scale.\\\\
 The interaction term between dark energy and dark matter is chosen of following three types \\
i)  proportional to the sum of the energy densities of the two dark components   \\
ii)  proportional to the product of the matter energy densities and \\
iii) proportional to dark energy density.\\
The  dynamical equations are reduced to an autonomous system for the three cases and corresponding phase space is analyzed.
\end{abstract}

\maketitle
%\pacs{Numbers 98.80.Cq, 95.36.+x}

%\keywords{Phantom dark energy field, Equilibrium point, Stability}

 %\maketitle

\section{Introduction}
In recent past, there were a series of  cosmological observations particularly from Type Ia Supernovae [1-3], Cosmic Microwave Background Radiation [CMBR][4], Baryon Accoustic Oscillation [5, 6] which  indicate cosmic acceleration. According to these observational evidences [7-9], at present  our universe consists of approximately $ 25 \% $ non-baryonic gravitating matter (known as dark matter (DM)), $ 70 \% $ non-baryonic non gravitating unknown matter (known as dark energy (DE)) and $ 5\% $ radiation and baryonic matter  which is well understood by the standard models of particles. This  dark matter is responsible for large scale structure formation in the universe.\\\\
 Although, the nature of the dark energy is still unknown, there are some possible candidates to describe it. A  natural choice for this DE  is the cosmological constant which represents a vacuum energy density having constant equation of state $ \omega = -1$ and this model ($\Lambda CDM $) fits well with large number of observations. But it suffers from two major drawbacks namely \textit{fine-tuning problem }( i.e why  the observed value is far below than the estimation from quantum field theory ?) and ii) \textit{coincidence problem }( why the constant vacuum energy and matter energy densities are precisely of the same order today?) As a result of these observational [10] and theoretical [11,12] probes for the cosmological constant, dynamical DE models have been proposed in the literature. Scalar field models [12 - 15] including  quintessence, K-essence, tachyon etc have attracted lots of  attention. Other models of dark energy include interacting DE models, braneworld models, chaplygin gas models and many others.\\
 Dominant contribution of the dark matter  and dark energy to the matter source of the universe draws a lot of interest in studying coupling between dark matter and dark energy [16 - 32].   Further, recently  it has been shown that proper choice of the  non-gravitational interaction term  may alleviate the coincidence problem [20 - 24].\\
 The holographic principle asserts that the number of relevant degrees of freedom of a system dominated by gravity must vary along with the area of the surface bounding the system [25].
 It is speculated in the literature that use of holographic principle [26, 27, 33] may shed some light into the unknown and mysterious nature of DE. Such a dark energy model is termed as Holographic Dark Energy (HDE) model. The HDE model is an attempt to apply holographic principle of the theory of quantum gravity to dark energy problem. Accordingly, the energy density of any given region should be bounded by that ascribed to a Schwarzschild black hole that fills the same volume [27, 33]. Mathematically, it means $ \rho_{d} \leq  M_{p}^{2}L^{-2}$, where $ \rho_{d}$ is the energy density for dark energy, $ M_{p}= (8\pi G)^{-\frac{1}{2}}$ is the reduced Planck mass and L is the size of the region (i.e IR cut off).

Usually from effective quantum field theory the DE density can be written as [ 26,33]
\begin{equation}
\rho_{d} = \frac{3M_{p}^{2}c^{2}}{L^{2}}
\end{equation}
where the dimensionless parameter 'c' is related to the uncertainties of the theory and factor 3 is for mathematical convenience. We can see that with L = constant, it portrays the cosmological constant model. \\
There are many choices for the IR cut off of which three are widely used in the literature.\\
  i)\textbf{ Hubble radius }i.e $ L = H^{-1}$  but it yields a wrong equation of state for dark energy but a correct DE density which is close to the observed value [21,26]. \\
  ii)\textbf{ Future event horizon}  i.e $ L = R_{E}$ : M Li suggested that future event horizon  should be chosen  as the IR cut off   to explain both the fine tuning and coincidence  problem [ 27, 32, 34 ].  \\
  iii) \textbf{Ricci's scalar curvature}  [28 - 31, 36] i.e $ L = (\dot{H} + 2H^{2})^{-\frac{1}{2}}$ : the reason for choosing this length scale is that it corresponds to the size of maximal perturbation, leading to the formation of black hole. This model avoids fine tuning problem and coincidence problem. People have studied extensively both HDE and interacting HDE models at different length scale [ 35, 36, 37, 38, 39 ].\\
  In the present work we first assume HDE model for  $ L = R_{E}$. Then we  consider modified holographic dark energy model   at Ricci's Scale (MHRDE)[40]. In both the cases, we have used interaction with  phenomenological form. We will study the models qualitatively and will check for viable cosmological solution considering cosmological constraints and observational data. In order to study  the dynamical character of the system, critical points are obtained. Any standard text will explain the method [41]. Corresponding cosmological models are analyzed. Feasible  cosmological solutions should depict our present universe as global attractor i.e all the possible initial conditions lead to the observed percentages of dark energy and dark matter, once reached, they remain fixed forever [42]. For this reason we will focus on the stability of critical points i.e cosmological models which are attractors. \\
 The paper is organized as follows: next section describes the basic equations and interaction terms of the model.   Section 3 presents a phase space analysis of  interacting HDE model taking future event horizon as IR cutoff  for two type of interactions.  Interacting MHRDE model has been considered in section 4 for phase space analysis and summary of the work has been presented in section 5.

\section{Basic Equations}
The homogeneous and isotropic flat FRW spacetime is chosen as the model of the universe. It is assumed to be filled up with DM in the form of dust (having energy density $\rho_{m}$) and HDE in the form of perfect fluid with variable equation of state i.e $p_{d}= \omega_{d}\rho_{d}$ with  $ \rho_{d},~p_{d}$ as the energy density and thermodynamic pressure of the dark energy respectively. Dark matter and dark energy interact non gravitationally with one another.

The Einstein field equations can be written as (in the units : $ 8\pi G = c = 1)$.
\begin{equation}
 3H^{2} = \rho_{m} + \rho_{d}
\end{equation}
\begin{equation}
   2\dot{H}  = - \rho_{m} - (1 + \omega_{d})\rho_{d}
\end{equation}

Using field equations (2) and (3) , the acceleration of the universe is given by
\begin{equation}
   \ddot{a} = - \frac{a}{6}[\rho_{m} + \rho_{d}(1 + 3\omega_{d})]
\end{equation}
which shows that  it is required to have $\omega_{d} < -\frac{1}{3}$ for cosmic acceleration. Introducing the density parameters, the first Friedman equation (i.e eqn (2)) can be written as
\begin{equation}
   \Omega_{m} +  \Omega_{d} = 1
\end{equation}
where
\begin{equation}
   \Omega_{m} = \frac{\rho_{m}}{3H^{2}} = \frac{u}{1 + u}~~~~and~~~~  \Omega_{d} = \frac{\rho_{d}}{3H^{2}}=\frac{1}{1 + u}
\end{equation}
with $ u = \frac{\rho_{m}}{\rho_{d}}$ , the ratio of energy densities.
  Now considering the interaction term between the two dark components the energy conservation relations take the form
\begin{equation}
\dot{\rho_{m}}+ 3H\rho_{m} = Q
\end{equation}
and
\begin{equation}
\dot{\rho_{d}}+ 3H(1 + \omega_{d})\rho_{d} = - Q
\end{equation}
Here Q, the interaction term is not unique.
 We choose $ Q > 0 $ so that there is transfer of energy from DE component to DM component i.e dark energy decays into dark matter. The positivity of Q ensures the validity of the second law of thermodynamics and satisfies the Le Chatelier's principle [16]. Also $ Q > 0 $ is in favour of resolving the coincidence problem. It should be noted that we have not included baryonic matter in the interaction due to the constraints imposed by local gravity measurements [16, 43]. In the next sections we shall consider three different choices of interaction term separately, namely (i) $ Q = 3b^{2}H\rho ~(\rho = \rho_{m} +\rho_{d},$ the total energy density), ii) $ Q = \frac{\nu}{H}\rho_{m}\rho_{d},(\nu > 0)$ and  iii)$ Q = 3 \nu H \rho_{d},~ (\nu > 0) $. The first choice of the interaction term is a particular case of a  general linear combination of the energy densities while the second choice is physically more viable in the sense that interaction rate vanishes if one of the energy densities is zero and increases with each of the densities.

\section{ Holographic DE model with event horizon as IR cut off }
The radius of the event horizon is defined by the improper integral
\begin{equation}
   R_{E} = a\int_{t}^{\infty}\frac{dt}{a}
\end{equation}
Note that above improper integral converges only when strong energy condition is violated. So in the present accelerating phase it always exists. Now choosing $ L = R_{E} $, we have from (1)
\begin{equation}
\rho_{d} = \frac{3M_{p}^{2}c^{2}}{R_{E}^{2}}
\end{equation}

From the conservation equations (7) and (8) and using the expression for energy density $\rho_{d} $ from equation (10), the equation of state parameter takes the  following form for any interaction Q,
 \begin{equation}
\omega_{d} = -\frac{1}{3}-\frac{2\sqrt{\Omega_{d}}}{3c}-\frac{Q}{3H\rho_{d}}
\end{equation}
% while the density parameter $\Omega_{d}$ evolves as
% \begin{equation}
%\dot{\Omega_{d}} = H{\Omega_{d}}^{2}( 1 - {\Omega_{d}})[\frac{1}{{\Omega_{d}}} + \frac{2\sqrt{{\Omega_{d}}}}{c}-\frac{3b^{2}}{(1-{\Omega_{d}}){\Omega_{d}}}]
%\end{equation}
%As a result the evolution of the ratio of the energy densities is given by
%\begin{equation}
%\dot{u}= H[-u (1 + \frac{2}{c\sqrt{(1 + u)} })+ 3b^{2}(1 + u)]
%\end{equation}
Now the second Friedmann equation (3) and the energy conservation for DE (i.e equation (8)) can be reduced (after a bit simplification) into an autonomous system as
\begin{equation}
\dot{H}= -\frac{3H^{2}}{2}[(1  -\frac{\Omega_{d}}{3}- \frac{2\Omega_{d}^{\frac{3}{2}}}{3c}- \frac{Q}{3H\rho}]
\end{equation}
\begin{equation}
\dot{\rho_{d}} = 2\rho_{d}H[\frac{\sqrt{\Omega_{d}}}{c}- 1]
\end{equation}
Now in $(\rho_{d}, H) $ phase plane this autonomous system has a line of critical points for $ \sqrt{\Omega_{d}} = c $ if
\begin{equation}
 Q = 3 ( 1 -c^{2})H \rho
\end{equation}
We shall now analyze the evolution equations for    different choices of the interaction term separately.
\subsection{  $ Q = 3b^{2}H\rho = 3b^{2}H(\rho_{m} +\rho_{d})$ }
  For the interaction, $ Q = 3b^{2}H\rho $,  the above autonomous system becomes
  \begin{equation}
\dot{H}= -\frac{1}{2}[3H^{2}(1 - b^{2})-\frac{\rho_{d}}{3}- \frac{2\rho_{d}^{\frac{3}{2}}}{3\sqrt{3}cH}]
\end{equation}
\begin{equation}
\dot{\rho_{d}} = 2\rho_{d}H[\frac{\sqrt{\Omega_{d}}}{c}- 1]
\end{equation}

    In the $ (\rho_{d}, H) $ phase plane the autonomous system (15) and (16)  has a line of critical points along the parabola
  $\rho_{d}= 3c^{2}H^{2}$, provided $ b^{2} = 1 - c^{2}$ from the above discussion. Critical points are given in table I. Direction field for the autonomous system given by equation (15) and (16) is drawn in Fig 1 for  c = 0.65. Miao Li et al studied constraints  on HDE model using Planck data [44]. The value of c is chosen here considering the constraints on 'c' from Planck data.  We can see that there is no stable critical point  i.e no attractor solution exists for the system. The determinant of linearized matrix for the autonomous system along  the parabola is $ - 8c^2H^2 < 0  $ representing the saddle type nature. The value of the cosmological parameters namely $\Omega_{d}= c^{2}$, $ \omega_{d}< -\frac{1}{c^{2}} < -1$ and $ u = \frac{b^{2}}{c^{2}}$ hold along the line of critical points. So we say that along the phase paths  the ratio of the energy densities bears a constant value partially solving coincidence problem and the universe will be in the phantom era representing accelerating universe. $E_{1} $ is one of those points lying on the line of critical points. But  these points are unstable.  There may be another critical point for $ \rho_{d}=0, ~$(0,~H), H unspecified if $ b^{2} = 1$. But the critical point is non-hyperbolic in nature, the determinant being zero. Therefore, no definite information about stability  can be extracted from this point, but it can describe a late time acceleration.
  \begin{figure}
 \includegraphics[width = 0.85000\linewidth]{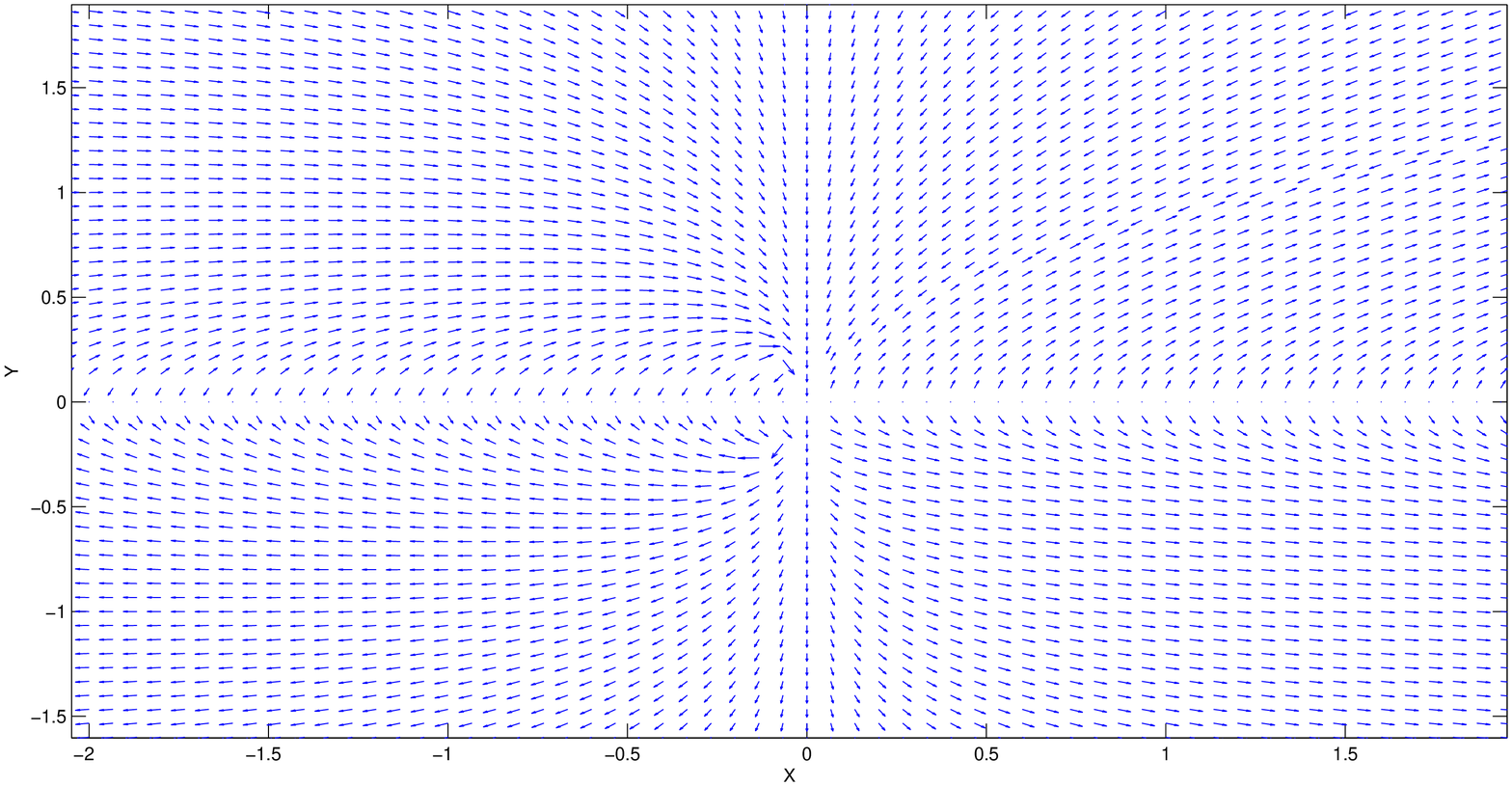}
\caption{\label{}Direction field for the autonomous system (15) and (16) for c = 0.65 }

 \end{figure}

  \begin{table}%[H] add [H] placement to break table across pages
\caption{\emph{Equilibrium points, their nature  for $  ~~ Q = 3b^{2}H\rho $}}
%$ E_{1}$ is any arbitrary point on the line of critical points $ \rho_{d}= 3c^2H^2 $
%\caption{}
 \begin{tabular}{|c|c|c|c|c|c|c|}
 \hline
 Equilibrium point & $\rho_{d}$ & H & $b^2$ & $  \Omega_{d} $ & $ \omega_{d}$   & Nature\\\hline

 $E_{1}$ &~ $3c^2H^2 $& ~H~ & $1- c^2$ &$c^2$ & $ <-1 $   & hyperbolic: saddle\\\hline
 $E_{2}$ &~ 0 & H~ & 1 ~& 0 & $< -\frac{1}{3}$ & non-hyperbolic\\\hline

 \end{tabular}

 \end{table}

  \begin{table}%[H] add [H] placement to break table across pages
\caption{\emph{Equilibrium points, their nature  for $  ~~ Q = 3b^{2}H\rho $}}
%\caption{}
 \begin{tabular}{|c|c|c|c|c|}
 \hline
 Equilibrium point & $\rho_{d}$ & $\rho_{m}$ &   $ \omega_{d}$   & Nature\\\hline

 $E_{1a}$ &~ $\rho_{d}$& $ \frac{b^{2}}{1 - b^{2}}\rho_{d}$ &   $ <-1 $   & non-hyperbolic \\\hline
 $E_{1b}$ &~ 0 & 0 &  undefined & undefined \\\hline

 \end{tabular}
 \end{table}

 Further the conservation equations (4) and (5) can be expressed in the form
 \begin{equation}
\dot{\rho_{m}} = \sqrt{ 3(\rho_{m}+\rho_{d})}[b^{2}\rho_{d}- (1 - b^{2})\rho_{m}]
\end{equation}
and
\begin{equation}
\dot{\rho_{d}} = -\sqrt{ 3(\rho_{m}+\rho_{d})}[b^{2}\rho_{m}+ (1 + \omega_{d}+ b^{2})\rho_{d}]
\end{equation}

Note that above set of equations (17) and (18) will form an autonomous set provided the equation of state parameter for DE is constant. For the autonomous system
 $ \dot\rho_{m} = 0$ and  $ \dot\rho_{d} = 0 $, we get   either  $ \rho_{m}=0, ~ \rho_{d}=0 $   or $ \rho_{m} =\frac{b^{2}}{1 - b^{2}}\rho_{d}  $ provided $ \omega_{d} = \frac{1}{ b^{2}- 1} < -1 $ which  supports cosmic acceleration.

    We see that from equation (18), in the phantom domain $ (\omega_{d} < -1 )$, $ \rho_{d} $ may begin to increase and dominate over dark matter. Thus the present model of the universe has dark energy dominance at early epoch and at late time accelerated scenario while dark matter dominates in the intermediate stages of the evolution.

%Moreover, u is less than unity in the early phase of evolution of the universe and then it increases gradually. Note that $ u \sim o(1) $ before $ \dot\rho_{m} = 0 $ or after $ \dot\rho_{m} = 0 $ or along the st. line $ \rho_{m} =\frac{b^{2}}{1 - b^{2}}\rho_{d} $ in the $ ( \rho_{m}, \rho_{d})$ plane provided $ b^{2} > or < or = \frac{1}{2} $.  Thus we may conclude that the coincidence problem is partially solved along the st line $ \rho_{m} =\frac{b^{2}}{1 - b^{2}}\rho_{d}  $ , but the present model can not explain for $ u \sim o(1) $ in the present scenario.

\subsection{ $ Q = \frac{\nu}{H} \rho_{m} \rho_{d},~ (\nu > 0 )$}

 For this choice of interaction term, the equation of state parameter is
\begin{equation}
\omega_{d} = -\frac{1}{3}-\frac{2\sqrt{\Omega_{d}}}{3c}-\nu (1- \Omega_{d})
\end{equation}
 and
 \begin{equation}
\dot{u}= 3Hu [-\frac{1}{3} - \frac{2}{3c \sqrt{(1 + u)}} + \frac{\nu }{1 + u}]
\end{equation}
and the evolution of density parameter is modified as
 \begin{equation}
\dot{\Omega_{d}} = H{\Omega_{d}}( 1 - {\Omega_{d}})[1 -3\nu\Omega_{d}+\frac{2\sqrt{\Omega_{d}}}{c}]
\end{equation}

 The second Friedmann equation (3) can be expressed in terms of density parameter as
\begin{equation}
\dot{H}= -\frac{3H^{2}}{2}[1 -\frac{\Omega_{d}}{3}- \frac{2\Omega_{d}^{\frac{3}{2}}}{3c}  - \nu\Omega_{d}(1 - \Omega_{d})]
\end{equation}
Here we have an autonomous system in the phase plane $( \Omega_{d}, H )$ corresponding to equations (21) and (22). In $(\Omega_{d}, H) $ phase plane this autonomous system has a line of critical points for $ \sqrt{\Omega_{d}} = c $   if either $ c^{2}= 1 $ or $ \nu c^2 = 1 $  and universe will be in phantom era. $E_{6}$  is one of those critical points. Fig 2 and Fig 3 presents the direction field  and phase portrait for this autonomous system for  c = .62 ( Note that  the whole  region  depicted in the figure is not  viable). The possible critical points are $E_{3}$ (H = 0, $ \Omega_{d}$ is unrestricted), $E_{4}$ (0, 0), $E_{5}$ (1, 0) and $E_{6}, ( c^{2}, \frac{1} {\nu c^{2}}, )$  if either $ c^{2}= 1 $ or $ \nu c^2 = 1 $. All the critical points are listed in table III. The critical points are all non hyperbolic. So stability of these points are not conclusive.

\begin{figure}
\begin{minipage}{.45\textwidth}
 \includegraphics[width = 1.000\linewidth]{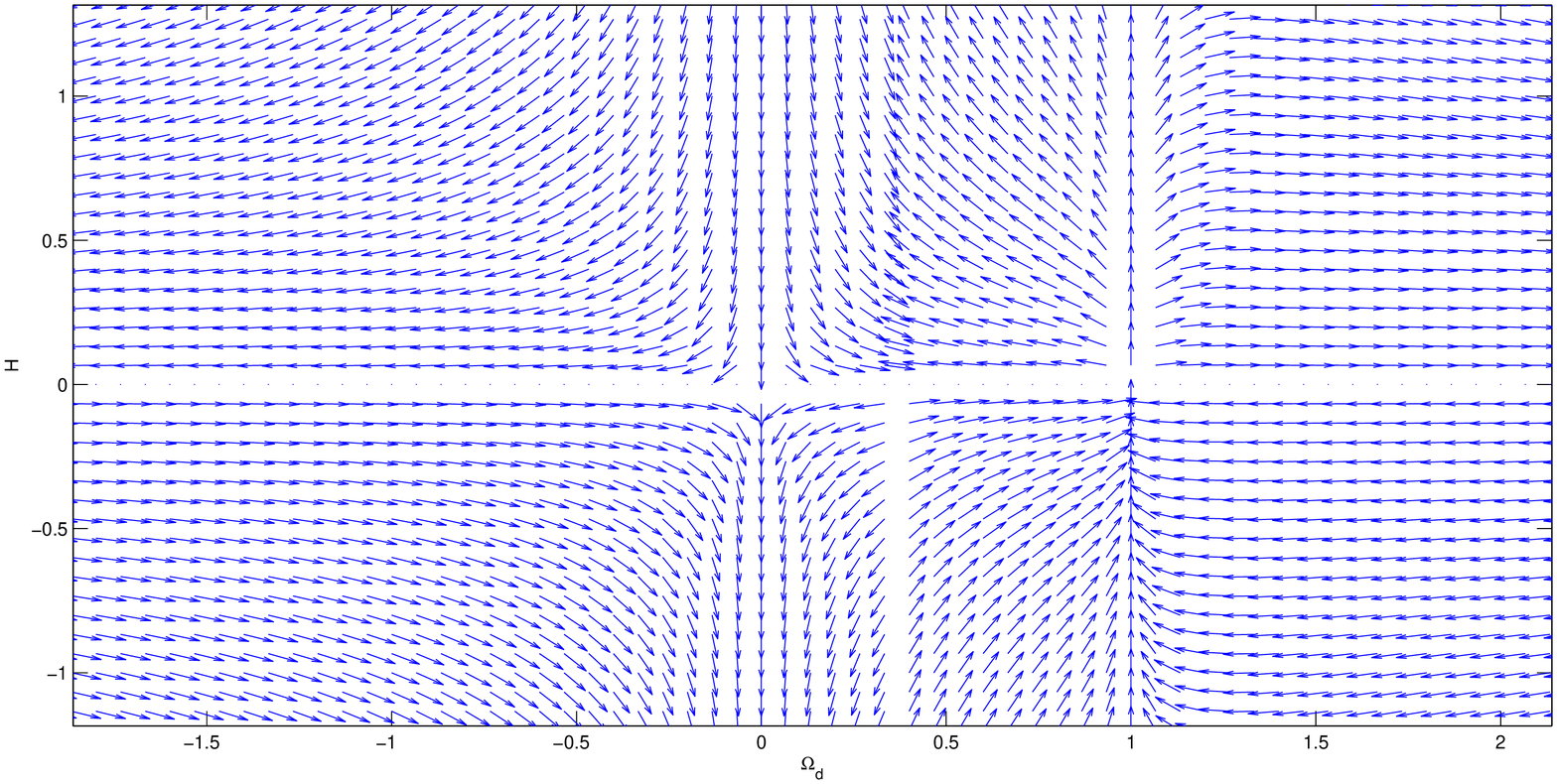}
\caption{\label{}direction field for the system (21)and (22)  for c = .62 }
 \end{minipage}
 %\begin{figure}
 \begin{minipage}{.45\textwidth}
 \includegraphics[width = 1.0\linewidth]{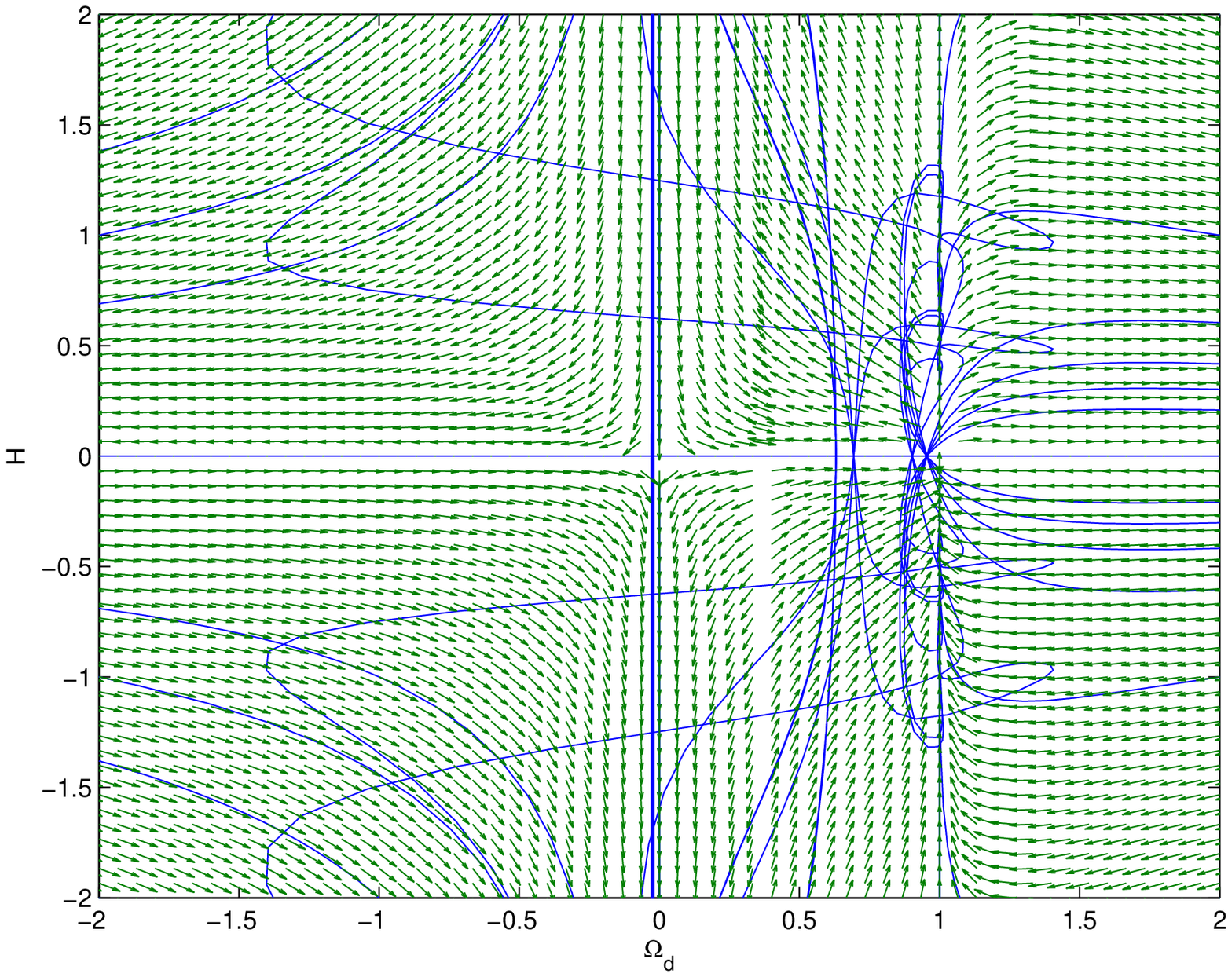}
\caption{\label{}phase portrait for the system (21)and (22) for c = .62 }
\end{minipage}
 \end{figure}
 \begin{table}%[H] add [H] placement to break table across pages
\caption{\emph{Equilibrium points, their nature  for $  Q = \frac{\nu}{H} \rho_{m} \rho_{d} $}}
%\caption{}
 \begin{tabular}{|c|c|c|c|c|c|}
 \hline
 Equilibrium point & $\Omega_{d}$ & H & $c^2$ &  $ \omega_{d}$   & Nature\\\hline

 $E_{3}$ &$\Omega_{d} $& ~0 &   & $ <-\frac{1}{3} $   & non-hyperbolic\\\hline
 $E_{4}$ &~ 0 & 0 &  ~&  $ -\frac{1}{3}-\nu$ & non-hyperbolic\\\hline
 $E_{5}$ &~1 & ~0 &  & $-\frac{1}{3}-\frac{2}{3c}$    &non- hyperbolic\\\hline
 $E_{6}$ &~ $c^2$ & $ \frac{1}{\nu c^2}$~ & 1 0r $\frac{1}{\nu}$~& $-1 + \nu c^2 -\nu$ & non-hyperbolic\\\hline

 \end{tabular}
 \end{table}

 In table III, the first three critical points with H = 0 represent static model of the universe. $E_{4}$ represents purely dark matter dominated  universe while critical point $E_{5}$  corresponds to universe filled with dark energy  only with $ \omega_{d} = -\frac{1}{3}- \frac{2}{3c}$   corresponding to phantom domain of  evolution of the universe. For the critical point $E_{6}$,  considering  $ \nu c^2 = 1 $ , $ u = \frac{1 - c^{2}}{c^{2}}$ and $ \omega_{d} = -\nu $. This critical point will be in the dark energy dominated era or dark matter dominated era according as $ c^{2}> ~or~c^{2} < \frac{1}{2}$. We can see that all the points describe late time acceleration. If $\nu > 1$, then $E_{6}$ crosses the phantom barrier. If $c^2 = 1$, this point is in phantom barrier. We are unable to establish it's stability, otherwise varying $\nu$  this critical point could describe present scenario of the universe.

Further the conservation equations (7) and (8) as before can be written as
\begin{equation}
\dot{\rho_{m}} = \rho_{m}[\frac{\nu}{H}\rho_{d} - \sqrt{ 3(\rho_{m}+\rho_{d})}]
\end{equation}
\begin{equation}
\dot{\rho_{d}} = -\rho_{d}[\frac{\nu}{H}\rho_{m} + ( 1 + \omega_{d}) \sqrt{ 3(\rho_{m}+\rho_{d})}]
\end{equation}
 Here $ \dot{\rho_{m}} = 0 $ along the curve $ \nu^{2}\rho_{d}^{2} =(\rho_{m} + \rho_{d})^2 $  or $ \nu \Omega_{d} = 1 $ and as a result both the energy components are of comparable magnitude i.e $ u \sim o(1)$ twice during the evolution and give a possible explanation to the coincidence problem.
\subsection{ $ Q = 3 \nu H \rho_{d} $}

In this case,
\begin{equation}
\dot{\Omega_{d}} = H{\Omega_{d}}( 1 - {\Omega_{d}})[1 - \frac{3\nu\Omega_{d}}{ 1 - \Omega_{d}} +\frac{2\sqrt{\Omega_{d}}}{c}]
\end{equation}
The second Friedmann equation (3) can be expressed in terms of density parameter as
\begin{equation}
\dot{H}= -\frac{3H^{2}}{2}[1 -\frac{\Omega_{d}}{3}- \frac{2\Omega_{d}^{\frac{3}{2}}}{3c}  - \nu\Omega_{d} ]
\end{equation}

 Like previous section, (25) and (26) form an autonomous system  in $(\Omega_{d}, H) $ phase plane . This autonomous system has a line of critical points   for $ \sqrt{\Omega_{d}} = c $   if  $ \nu = \frac{1- c^2}{c^2 }$. $E_{7}$  represents one of those critical points. These  critical points will be in the dark energy  dominated era or dark matter dominated era according as $ c^{2}> ~or~c^{2} < \frac{1}{2}$. We find that there are  other critical points also namely, $E_{8}$  $(\Omega_{d}, 0)$, $\Omega_{d}$ unrestricted, $E_{9}$ (1, 0), $E_{10}$ (1, H) for $ \nu =\frac{2}{3} ( 1 - \frac{1}{c}) $ , $ H \neq 0 $( see table IV). The points describe late time acceleration, but all of them are non-hyperbolic critical points so could not infer the stability of these points.
 \begin{table}%[H] add [H] placement to break table across pages
\caption{\emph{Equilibrium points, their nature  for $   Q = 3 \nu H \rho_{d}, \nu > 0  $}}
%\caption{}
 \begin{tabular}{|c|c|c|c|c|c|}
 \hline
 Equilibrium point & $\Omega_{d}$ & H & $\nu$ &  $ \omega_{d}$   & Nature\\\hline

 $E_{7}$ &$c^2 $& ~H & $\frac{1-c^2}{c^2}$  & $ -\frac{1}{c^2} $   & non-hyperbolic\\\hline
 $E_{8}$ & $\Omega_{d}$ & 0 &  ~&  $ <-\frac{1}{3}-\nu$ & non-hyperbolic\\\hline
 $E_{9}$ & 1 & 0 &  ~&  $ <-\frac{1}{3}- \frac{2}{3c}-\nu$ & non-hyperbolic\\\hline

 $E_{10}$ &~1 & ~H($\neq 0) $ & $ \frac{2}{3}(1- \frac{1}{c})$ & -1    &non- hyperbolic\\\hline

 \end{tabular}
 \end{table}
 Like the previous section the  critical points with H = 0 i.e $E_{8}$  and $E_{9}$  represent static model of the universe. $E_{10}$ is interesting in the sense that it represents  accelerating universe purely dominated by dark energy.\\
 For  the above three interactions  sometimes we  could not conclude on stability of critical points being non-hyperbolic, though this model solves coincidence problem partially.

\section{  Modified Holographic  Ricci Dark Energy  Model  }
Here we associate infrared cutoff $ L $ with dark energy density which can be written as [45 ]
\begin{equation}
       \rho_{d} = \frac{2}{\alpha - \beta}(\dot{H} + \frac{3\alpha}{2}H^{2})
\end{equation}

     $ \alpha, \beta $ being free constants. We have chosen $\alpha, \beta$ remembering the range preferred by authors [46,47] estimating the  parameters in most cases. But to have a clear understanding, we have used several values in plotting phase portraits. The whole phase space may not be the feasible region considering cosmological constraints.
Also the equation of state parameter for DE takes the  form
\begin{equation}
\omega_{d} = - (\alpha - \beta) + \frac{\alpha}{\Omega_{d}} - \frac{1}{\Omega_{d}}
\end{equation}
Using field equations , the deceleration parameter is given by
\begin{equation}
q = - \frac{\ddot{a}}{ 2H^{2}} =  \frac{3}{2}\alpha -  1 - \frac{3}{2(1 + u)}(\alpha - \beta)
\end{equation}
 We shall now analyze the dynamical equations for the three prescribed choices of the interactions.

\begin{figure}
\begin{minipage}{.45\textwidth}
\includegraphics[width = 1.0\linewidth]{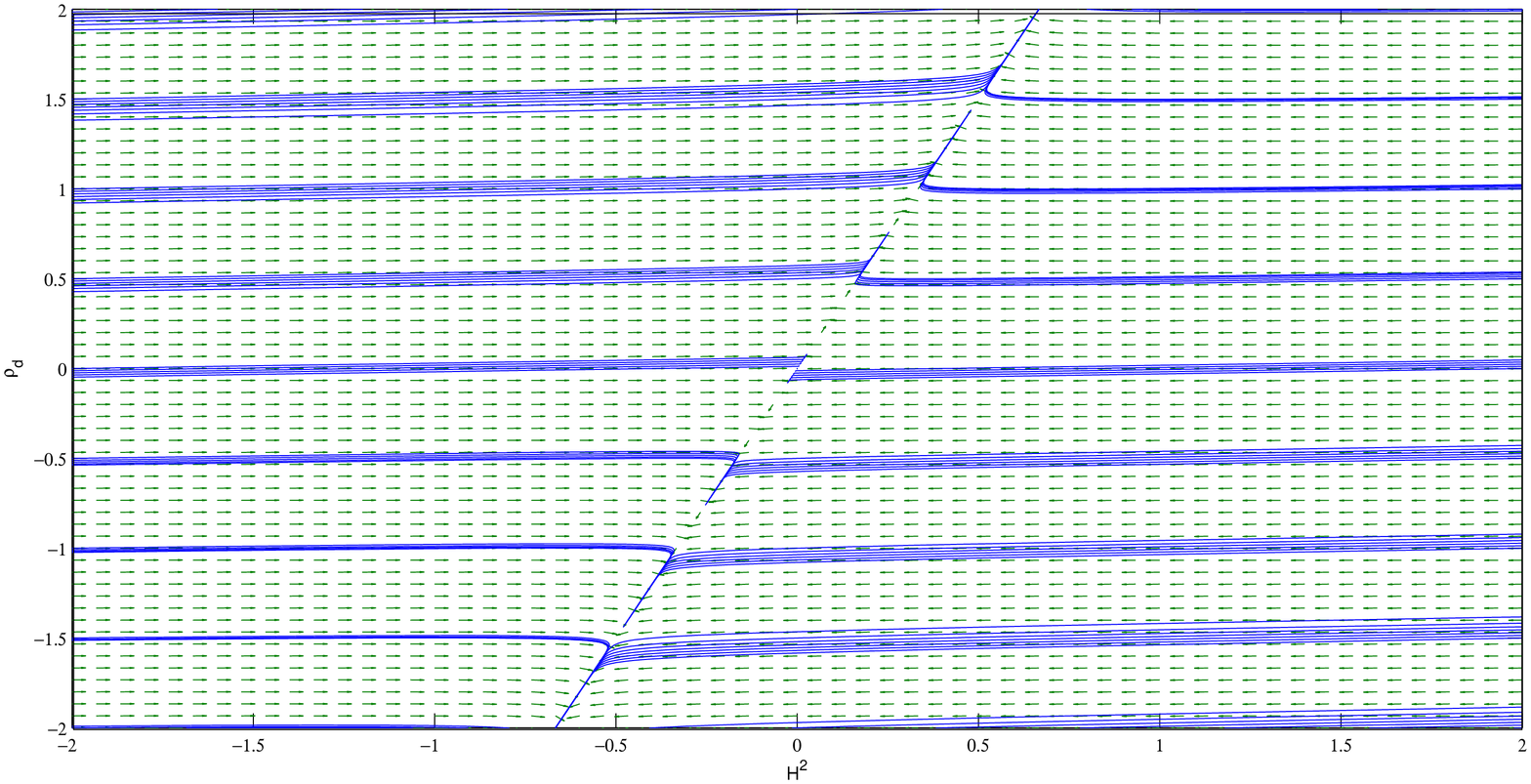}
\caption{\label{}phase portrait for the system (34)and (35) where $ \alpha = 1.01 , \beta = -0.01 , b^{2} = .01 $ }
 \end{minipage}
 %\end{figure}
\begin{minipage}{.45\textwidth}
\includegraphics[width = 1.000\linewidth]{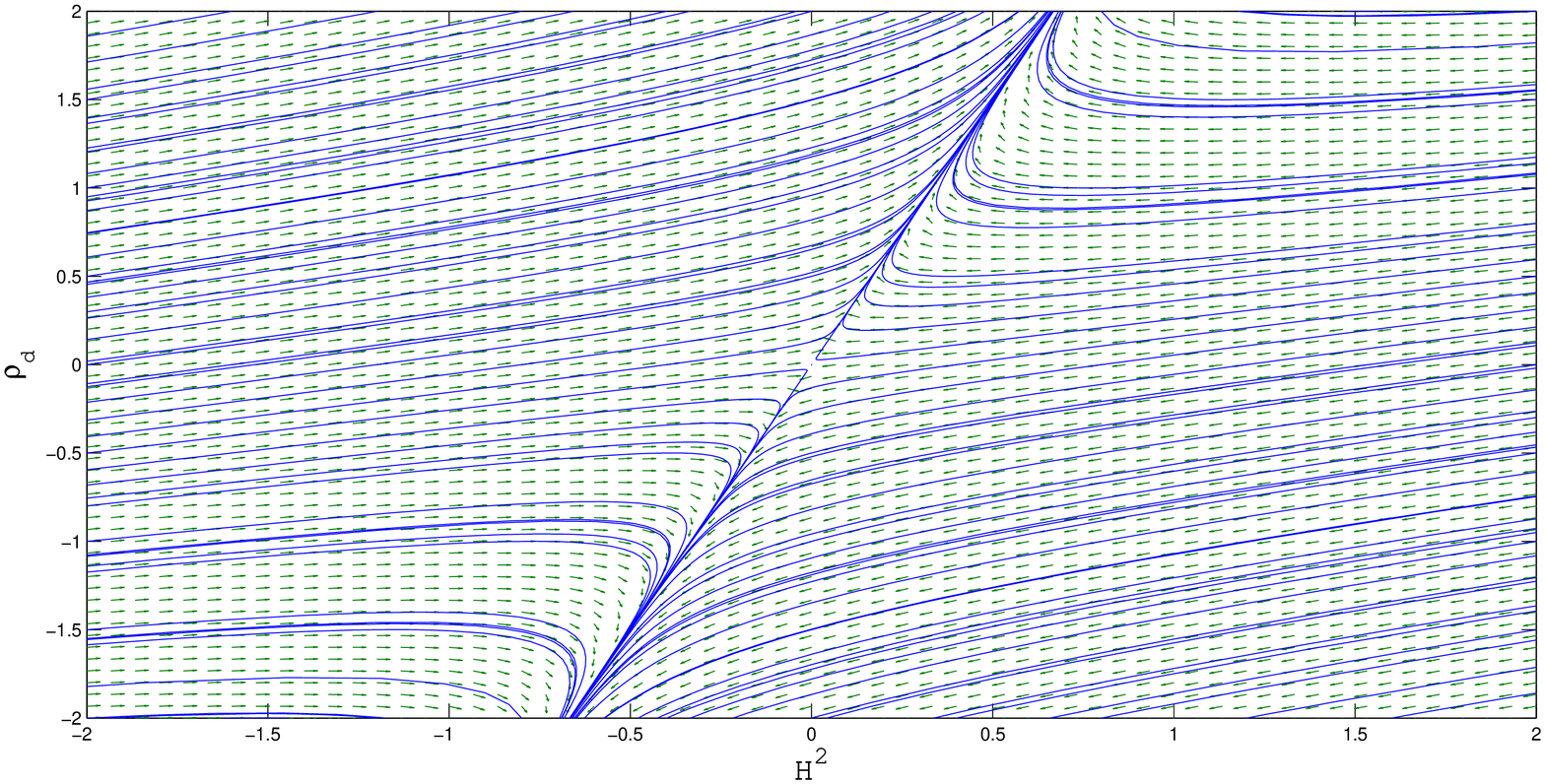}
\caption{\label{}phase portrait  for the system (34)and (35) where $ \alpha = 1.1, \beta = -.13, b = .01  $  }
\end{minipage}
 \end{figure}

\subsection{ $ Q = 3 b^{2}H\rho$ }
Using field equations and the above form of the equation of state parameter, the energy conservation equations can be written explicitly in the form
 \begin{equation}
\dot{\rho_{m}} = \sqrt{ 3(\rho_{m}+\rho_{d})}[b^{2}\rho_{d}- (1 - b^{2})\rho_{m}]
\end{equation}
and
\begin{equation}
\dot{\rho_{d}} = -\sqrt{3(\rho_{m} + \rho_{d})}[ b^{2}(\rho_{m}+ \rho_{d}) + ( 1 + \omega_{d}) \rho_{d}]
\end{equation}

So the evolution of the density parameter and u, the ratio of the energy densities are given by
\begin{equation}
\dot{\Omega_{d}} = -3H[-\Omega_{d}(1-\Omega_{d})(\alpha - \beta)+ (\alpha - 1)( 1 - {\Omega_{d}})+ b^{2}]
\end{equation}
and
\begin{equation}
\dot{u} = 3H[-(\alpha - \beta)u + u (1 + u )(\alpha - 1 ) + b^{2}(1 + u)^{2}]
\end{equation}

Equation (31) shows that $ \rho_{d}$ gradually decreases with the evolution of the universe till the universe enters the phantom era ( i.e $ \omega_{d} < -1 $ ). Thus if DE density is assumed to be sufficiently large at early phases of the evolution then from equation (30) $ \rho_{m}$ increases till some intermediate stage when $ \dot{\rho_{m}} = 0 $ and then decreases. Here the transition of $ \rho_{m}$ occurs along the straight line
$ \rho_{m} =\frac{b^{2}}{1 - b^{2}}\rho_{d}  $ in the $ ( \rho_{m}, \rho_{d})$ plane. So similar to the previous section , it may be a possible resolution of the coincidence problem  and the late time dominance by DE matches with present day observations.
Moreover, choosing  x = ln a as the time variable, the second Friedmann equation (3) and the conservation  equation (8) can be formulated as a linear homogeneous autonomous system in the phase plane $ (\rho_{d}, H^{2})$ :
\begin{equation}
(H^{2})' = -3\alpha H^{2} + (\alpha - \beta)\rho_{d}
\end{equation}
\begin{equation}
\rho_{d}' = -9( \alpha -1 + b^{2})H^{2} - 3\rho_{d}(1 - \alpha + \beta)
\end{equation}
Here $ (0,0)$ i.e origin is the critical point (see table V). The determinant of the linearized matrix is $9(\alpha - \beta)b^2 + 9\beta $.  If $ b^2 < \frac{\beta}{\beta-\alpha}$ the critical point is a saddle  point and if  $ b^2 > \frac{\beta}{\beta-\alpha}$  then it is stable but it can not be a viable cosmological solution. Nature of this  critical point is presented in Fig 4 and Fig 5 considering some values for $ \alpha $ and $ \beta$. Fig 4 is drawn choosing the values of $\alpha, \beta $ as suggested by the authors [45] for this interaction  considering the observational support.

\begin{table}%[H] add [H] placement to break table across pages
\caption{\emph{Equilibrium points, their nature  for $  ~~ Q = 3b^{2}H\rho $}}
%\caption{}
 \begin{tabular}{|c|c|c|c|c|}
 \hline
 Equilibrium point & $H^2$ & $ \rho_{d}$ &   $ q $   & Nature\\\hline

 $E_{1a}$ &~ 0 & 0 &   $ \frac{3\beta}{2}-1 $   & hyperbolic: saddle if $b^2<\frac{\beta}{\beta-\alpha}$, stable if $b^2>\frac{\beta}{\beta-\alpha}$\\\hline

 \end{tabular}
 \end{table}

 \subsection{ $ Q = \frac{\nu}{H} \rho_{m}\rho_{d}, \nu > 0$}
The explicit form of the energy conservation equations are given by
\begin{equation}
\dot{\rho_{m}} = \rho_{m}(\frac{\nu}{H}\rho_{d} - 3H)
\end{equation}
and
\begin{equation}
\dot{\rho_{d}} = -\rho_{d}[\frac{\nu}{H}\rho_{m} + 3H( 1 + \omega_{d}) ]
\end{equation}

As a result the evolution of u ( ratio of energy densities) obeys
\begin{equation}
\dot{u} = 3Hu [( \alpha -1)(1 + u) - (\alpha - \beta)  + \nu  ]
\end{equation}
\begin{figure}
 \begin{minipage}{.45\textwidth}
 \includegraphics[width = 1.0\linewidth]{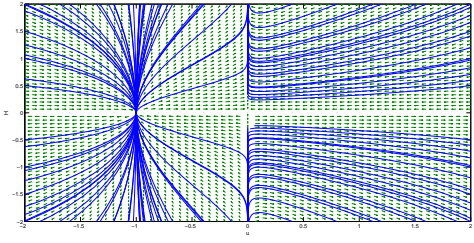}
  \caption{\label{} phase portrait for the system (38)and (39) where $ \alpha = 1.01 , \beta = -0.01, \nu = .1 $ }
 \end{minipage}
 %\end{figure}
 \begin{minipage}{.45\textwidth}
%\begin{figure}
 \includegraphics[width = 1.00\linewidth]{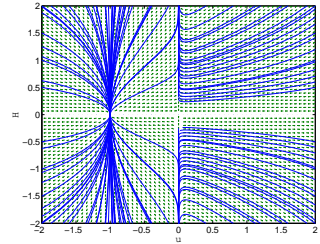}
 %\caption{\label{}  }
\caption{\label{} Phase portrait for the system (38) and (39) where $ \alpha = 1.1 , \beta = -.13, \nu = .1 $ }
\end{minipage}
 \end{figure}
 The second Friedmann equation (3) becomes
\begin{equation}
\dot{H}= -\frac{3H^{2}}{2}[\alpha- \frac{\alpha - \beta}{(1 + u)}]
\end{equation}
The equations (38) and (39) form an autonomous system in (u, H) plane. Phase portrait of this system is drawn in Fig 6 and Fig 7 for some $\alpha, \beta $.  There is  a line of critical points for H = 0 but not of physical interest.  Another critical point  is $ (\frac{-\beta}{\alpha}$,  H ) where H is unspecified if $\alpha\nu + \beta - 1 = 0 $. These points are listed in table VI. Again  the  critical points are non hyperbolic. From the figure we can see that no stable point exists. If $\beta $ is negative, u is valid constant i.e energy densities of dark matter and dark energy somewhat comparable, then $E_{4b}$ having present  acceleration can be interesting.

\begin{table}%[H] add [H] placement to break table across pages
\caption{\emph{Equilibrium points, their nature  for $  Q = \frac{\nu}{H} \rho_{m} \rho_{d} $}}
%\caption{}
 \begin{tabular}{|c|c|c|c|c|c|}
 \hline
 Equilibrium point & $ u $ & H & $\nu$ &  $ q $   & Nature\\\hline

 $E_{2b}$ &$ u $& ~0 &   & $\frac{3\alpha}{2}-1 - \frac{3(\alpha-\beta)}{2(1 + u)}$   & non-hyperbolic\\\hline
 $E_{3b}$ &~ 0 & 0 &  ~&  $\frac{3\beta}{2}-1 $ & non-hyperbolic\\\hline
 $E_{4b}$ &$\frac{-\beta}{\alpha}$ & H & $ \frac{1 - \beta}{\alpha}$ &  -1   &non- hyperbolic\\\hline

 \end{tabular}
\end{table}

\subsection{ $ Q = 3 \nu H \rho_{d}, \nu > 0 $}

 In this case we consider  the autonomous system in (u, H)  plane like previous section
 \begin{equation}
\dot{u} = 3Hu [( \alpha -1)u - 1 + \beta)  + \nu  + \frac{\nu}{u} ]
\end{equation}

\begin{equation}
\dot{H}= -\frac{3H^{2}}{2}[\alpha- \frac{\alpha - \beta}{(1 + u)}]
\end{equation}

\begin{table}%[H] add [H] placement to break table across pages
\caption{\emph{Equilibrium points, their nature  for $  Q = 3\nu H  \rho_{d} $}}
%\caption{}
 \begin{tabular}{|c|c|c|c|c|c|}
 \hline
 Equilibrium point &  u  & H & $ \nu$ &   q   & Nature\\\hline

 $E_{5b}$ & u & ~0 &   & $\frac{3\alpha}{2}-1 - \frac{3(\alpha-\beta)}{2(1 + u)}$    & non-hyperbolic\\\hline
 $E_{6b}$ &~ 0 & 0 &  ~&  $\frac{3 \beta}{2}-1 $ & non-hyperbolic\\\hline
 $E_{7b}$ & $\frac{-\beta}{\alpha}$ & H & $ -\frac{  \beta}{\alpha}$ &  -1   & non- hyperbolic\\\hline

 \end{tabular}
\end{table}
 Phase portrait of this system  consisting of (40) and (41) is drawn in Fig 8 and Fig 9. No stable attractor exists in the figure. Again, there is  a line of critical points for H = 0 but not of physical interest. Another critical point may be $ (-\frac{\beta}{\alpha}, H )$  provided $ \nu = -\frac{\beta}{\alpha} $ ( see table VII). We have taken  negative value of $\beta $ here to have valid $\nu$ for the point $E_{7b}$. These points are non hyperbolic in nature so local stability  is not conclusive.
 \begin{figure}
\begin{minipage}{.45\textwidth}
\includegraphics[width = 1.0\linewidth]{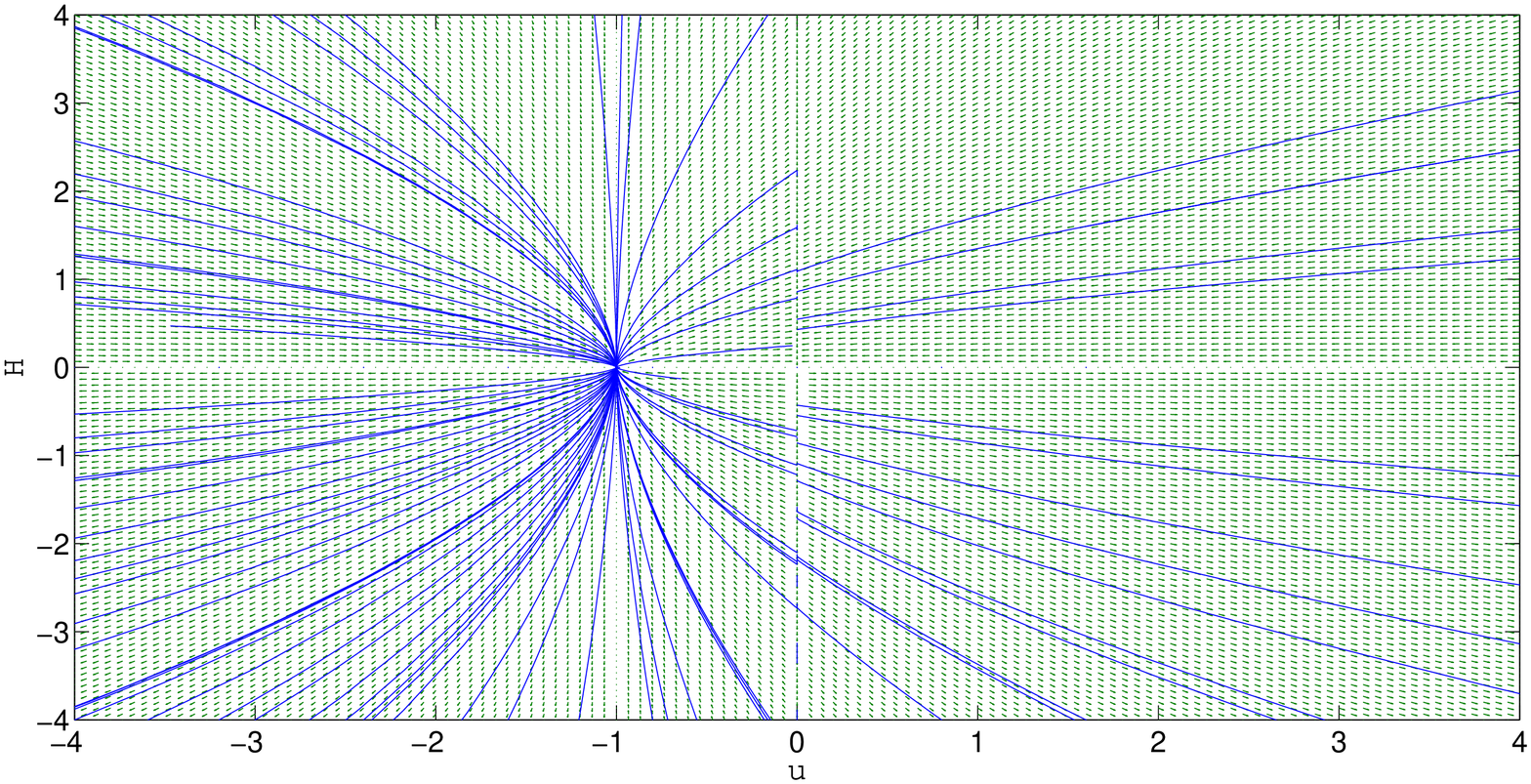}
\caption{\label{}phase portrait for the system (40)and (41) where $ \alpha = 1.01 , \beta = -0.01 , \nu = .001 $ }
 \end{minipage}
 %\end{figure}
\begin{minipage}{.45\textwidth}
\includegraphics[width = 1.000\linewidth]{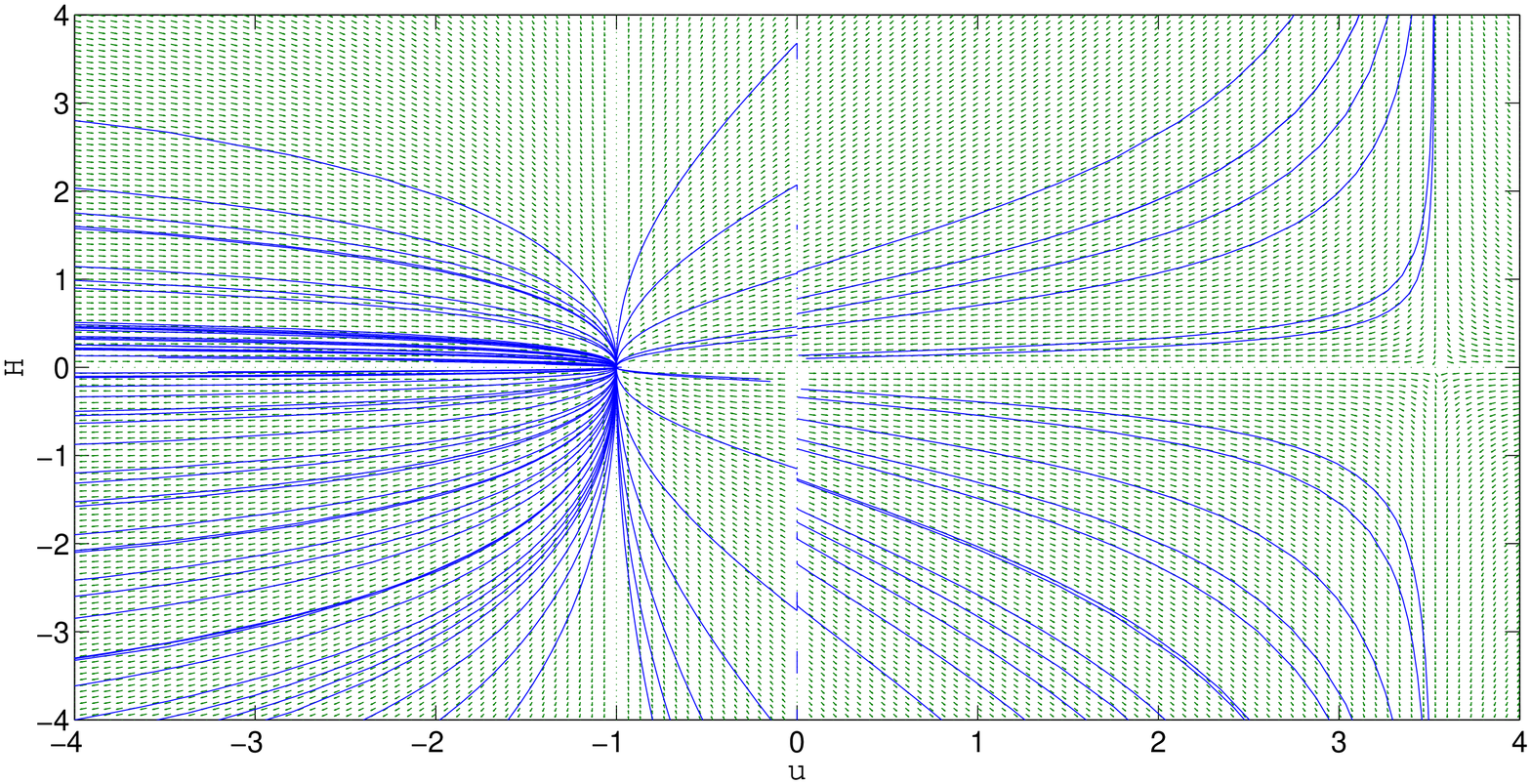}
\caption{\label{}phase portrait  for the system (40)and (41) where $ \alpha = 1.3 , \beta = -0.1, \nu = .001  $  }
\end{minipage}
 \end{figure}

\section{ summary}
The present work deals with the wellknown interacting holographic dark energy model from the point of view of dynamical system analysis. The cosmic fluid consists of  two interacting dark sectors namely dark matter in the form of dust and (modified) holographic  dark energy in the form of perfect fluid. The energy density and  the variable equation of state parameter for the dark energy is determined from the holographic principle. The interaction  terms between the two  dark sectors are chosen phenomenologically but are widely used in literature. Here we have considered three choices of the interaction term to avoid mathematical complexity - i) proportional to the total energy density of the dark sectors, ii) proportional to the product of energy densities of dark matter and  dark energy and iii) proportional to dark energy  density. However, we have not considered interaction with baryonic matter due to constraints imposed by local gravity measurements. The future event horizon   is chosen as IR cut off for holographic dark energy model   while IR cut off for modified holographic dark energy is chosen  at Ricci length scale.  In both the models, with three choices of interaction, the evolution equations are suitably  reduced to an autonomous system. In most of the cases, we have found line of critical points which are presented in tables. The nature of the hyperbolic   critical points and their stability are analyzed using jacobian matrix of the dynamical system.\\
We can see from the tables that for the HDE model,  three different interacting terms lead to same line of critical points $ \Omega_{d}= c^2$  with different constraints on $\nu $ and c. An arbitrary point on this curve can cross the phantom barrier depending on values of $\nu $ and c. The points $E_{6}$, $E_{7}$ are among those points. For MHRDE model $E_{4b}$, $E_{7b}$  are interesting as they describe accelerating universe. But we could not conclude on  stability of  the above points due to their non-hyperbolic nature.

\begin{acknowledgments}
The authors are thankful to IUCAA, Pune for warm hospitality ,  nice research environment and facilities at the library as  major part of the  work is done  during a visit to IUCAA. The authors are also thankful to UGC-DRS programme, Department of Mathematics, Jadavpur University.
\end{acknowledgments}
\section{References}

\end{document}